\documentclass[superscriptaddress]{revtex4}

\usepackage{amsmath,amsfonts,amssymb}
\usepackage{xcolor}
\usepackage[bookmarks=false]{hyperref}
\numberwithin{equation}{section}

\newcommand{\tr}[1]{\left(#1\right)}

\def\be{\begin{equation}}
\def\ee{\end{equation}}
\def\ba{\begin{array}}
\def\ea{\end{array}}
\def\bea{\begin{eqnarray}}
\def\eea{\end{eqnarray}}

\def\ical{\mathcal{I}}

\def\sfrac#1#2{{\textstyle\frac{#1}{#2}}}
\def\sbinom#1#2{{\textstyle\binom{#1}{#2}}}
\def\part#1{\frac{\partial}{\partial#1}}

\def\C{\mathbb C}
\def\R{\mathbb R}
\def\NN{\mathbb N}

\begin{document}

\begin{flushright}
ITP-UH-02/14
\end{flushright}

\title{The structure of invariants in conformal mechanics}

\author{Tigran Hakobyan}
\email{hakob@ysu.am}
\affiliation{Yerevan State University, 1 Alex Manoogian, 0025 Yerevan, Armenia}
\author{David Karakhanyan}
\email{karakhan@yerphi.am}
\affiliation{Yerevan Physics Institute, 2 Alikhanyan br., 0036 Yerevan, Armenia}
\author{Olaf Lechtenfeld}
\email{lechtenf@itp.uni-hannover.de}
\affiliation{Leibniz Universit\"at Hannover, Institut f\"ur
Theoretische Physik, Appelstr.\ 2, 30167 Hannover Germany}

\begin{abstract}
We investigate the integrals of motion of general conformal mechanical systems
with and without confining harmonic potential as well as of the related angular subsystems,
by employing the SL(2,$\R$) algebra and its representations. In particular, via the
tensor product of two representations we construct new integrals of motion from old ones.
Furthermore, the temporally periodic observables (including the integrals) of the angular subsystem  
are explicitly related to those of the full system in a confining harmonic potential. 
The techniques are illustrated for the rational Calogero models and their angular subsystems, 
where they generalize known methods for obtaining conserved charges beyond the Liouville ones.
\end{abstract}
\maketitle

\section{Introduction}

\noindent
Arguably the most important one-dimensional multi-particle system
is defined by the inverse-square two-body interaction potential.
It has been introduced by Calogero four decades ago
and is integrable both with and without a confining harmonic potential
\cite{calogero69,moser}:
\begin{gather}
\label{Calogero}
H_\omega= \frac{1}{2}\sum_{i=1}^N \left( p_i^2 + \omega^2 q_i^2\right)
     +\sum_{i<j}\frac{g^2}{(q_i-q_j)^2}\ ,
\\
\label{Calogero-0}
H_0=\frac{1}{2}\sum_{i=1}^N p_i^2 + \sum_{i<j}\frac{g^2}{(q_i-q_j)^2}\ .
\end{gather}
This system continues to attract much interest due to its rich internal structure and
numerous applications. So far, various integrable
extensions have been constructed and studied,
in particular, for trigonometric potentials \cite{trig-Cal},
for particles with spins \cite{spin-Cal},
for supersymmetric systems \cite{super-Cal}, and for other Lie algebras \cite{algebra}.
Such systems exhibit a rich spectrum of physical properties:
fractional statistics \cite{frac-stat}, Laughlin-type wavefunctions \cite{frac-Hall}
and a resonance valence-bond ground state for a related spin chain \cite{haldane-rvb}.
Calogero models appear in many areas of physics and mathematics,
like black holes \cite{gibbons}, quantum hydrodynamics, or orthogonal polynomials.

The rational Calogero models are maximally superintegrable,
i.e.\ they possess $N{-}1$ additional integrals of motion.  For the classical Hamiltonian without
confining harmonic potential, these have been constructed explicitly by Wojciechowski \cite{woj83}.
Later this construction was extended to the quantum case \cite{kuznetsov,gonera98-2}
and to the inclusion of a confining harmonic potential, where oscillatory behavior with
commensurate frequencies implies the superintegrability \cite{adler,perelomov81,gonera99-2}.
This property has been established also for the hyperbolic Calogero model \cite{gonera98-1} and
the relativistic extension known as the rational Ruijsenaars-Schneider model \cite{feher10,feher12}.

An important feature of rational Calogero models
is the dynamical conformal  SL(2,$\R$) symmetry,
\be
\label{sl2}
\{H_0,D\}=2H_0,\qquad \{K,D\}=-2K,\qquad \{H_0,K\}=D,
\ee
generated  by  the Hamiltonian (\ref{Calogero-0})
together  with the dilatation  and conformal boost
generators \cite{woj-sl2}
\be
\label{DK}
D=\sum_{i=1}^N p_iq_i, \qquad
K= \frac12\sum_{i=1}^N q^2_i.
\ee
Many properties of these systems, like superintegrability \cite{gonera98-1,gonera99-2},
equivalence to a free-particle system \cite{gonera99-1,lech06}, or the existence of action-angle
variables \cite{gonera00} are simple consequences of the conformal symmetry.
The Casimir element of this algebra,
\be
\label{ical}
\ical =4H_0 K - D^2\ ,
\ee
coincides with the angular part of the Calogero model and is
an integral of motion of both Hamiltonians
\eqref{Calogero} and \eqref{Calogero-0}.
It does not belong to the usual system
of Liouville integrals, but its commutator
with them produces all additional integrals
of motion responsible for the superintegrability.
The angular part \eqref{ical} can be considered
as a separate (super)integrable system describing a
particle moving on the $(N{-}1)$-dimensional sphere, which
has been defined and studied in a number of recent papers
\cite{cuboct,hkln,hlns,hln,flp}.

The rational Calogero models are integrable members of the more
general class of {\it conformal\/} mechanical systems, whose action is
invariant under the conformal transformations \eqref{sl2}
and which were first introduced in \cite{conf-mech}.
As a recent application, such systems can describe particle dynamics near the
horizon of an extremal black hole \cite{kumar98,gal-ners13,galaj-bh}.
A lot of what is derived in this paper also applies to any conformal mechanics system.
The article is structured in the following way.

In Section 2, we describe the SL(2,$\R$) representation content of integrals of motion
in conformal classical mechanics and employ the conformal algebra to expand the system
of conserved charges.
Any such integral of motion with a definite value of the conformal spin is, by definition,
a highest-weight state and thus generates a representation of the conformal algebra.
The descendant states in this representation are not conserved, but can be used to
construct additional integrals \cite{woj83} which, of course, are highest weights
in another representation.
The tensor product of two representations is a convenient way to generate new ones.
Thus, given two integrals of conformal classical mechanics, we may decompose
the tensor product of their conformal representations into irreducible pieces
and pick from each of these the highest state, which will yield a new conserved charge.
In this way, extending a method applied in \cite{woj83,gonera98-2},
we express new integrals of conformal classical mechanics in terms of descendants
of two given integrals of motion.
The simplest application to the rational Calogero model recovers Wojciechowski's
construction \cite{woj83}.
From the standard tensor product, the descendant states are combined symmetrically,
i.e. the corresponding phase-space functions are pointwise multiplied.
When instead one combines them under the Poisson bracket,
it seems that merely the known Liouville integrals are reproduced, but
this option needs a more detailed study.

In Section 3, the aforementioned construction of additional integrals of motion
is extended to the quantum case.  The new integrals appear in symmetrized products of
operator-valued conformal representations, which in the semiclassical limit reduce
to pointwise products.

Section 4 is devoted to the integrals of motion
for conformal mechanics in a confined harmonic potential, e.g.~\eqref{Calogero},
and to those for the related angular mechanics \eqref{ical}.
Recently, it has been shown that the spectrum and eigenstates of
these two systems are closely related for the quantum Calogero model \cite{flp}.
Here we study this relation at the level of more general conformal mechanical systems.
Starting with one or more integrals of motion for some unconfined conformal mechanics,
e.g.~\eqref{Calogero-0}, a system of oscillating observables is constructed for the
corresponding angular mechanics. The frequencies of these observables are integer
multiples of the basic frequency, the latter being the square root $\sqrt{\ical}$ of the
angular Hamiltonian \cite{hkln,hlns}. These observables can easily be combined
to products with vanishing frequency, giving rise to further integrals of motion.
Similarly, for the model with a confining harmonic potential, oscillating observables
and integrals of motion are derived from conserved charges of the unconfined system,
but now the basic frequency is the frequency $\omega$ of the confining potential
\cite{gonera98-2}. In both cases we employ appropriate SL(2,$\R$) rotation operators.
For the angular system, in addition a noncanonical special conformal transformation
inverting the radial coordinate is involved. As a result, we have found the exact
relation between the oscillating observables of confined conformal mechanics and of
the related angular system.

Section 5 revisits the matrix-model construction of the Calogero Hamiltonians $H_0$ and
$H_\omega$. We consider the additional integrals described in Section~2 and generated from
the standard Liouville integrals of the Calogero Hamiltonian $H_0$. Again based on these
Liouville integrals, we construct the oscillating observables of the confined system
$H_\omega$ treated in Section~4 and describe them in terms of oscillating matrices.
We then prove that the Poisson action of the angular Hamiltonian $\ical$ on the standard
Liouville integrals of $H_\omega$ produces $N{-}1$ additional integrals, which combine with the
$N$ Liouville integrals to a complete and independent system.
This generalizes a similar property for the unconfined Calogero Hamiltonian $H_0$
\cite{hkln,hlns}.

\newpage

\section{SL(2,$\R$) structure of the integrals of motion: classical case}

\noindent
For any function $f$ on phase space, define the associated
Hamiltonian vector field by the Poisson bracket action
\be
\label{hat-f}
\hat f=\{f,\cdot\}.
\ee
The assignment $f\to\hat f$
is a Lie algebra homomorphism, the constants on the phase space form its kernel.
For an interaction potential~$V$, the vector fields
\be
\label{hat-sl2}
\hat H_0=\sum_i \left(p_i\part{q_i} - \frac{\partial V}{\partial q_i}\part{p_i}\right),
\qquad
\hat K=-\sum_i q_i\part{p_i},
\qquad
\hat D=\sum_i \left(q_i\part{q_i}-p_i\part{p_i}\right)
\ee
satisfy the $sl(2,\R)$ algebra \eqref{sl2},
and the vector field of the Casimir element  $\hat\ical$, of course, commutes with them.

Consider now the general conformal mechanical system with the Hamiltonian $H_0$ obeying the
symmetry relations \eqref{sl2}.
First, this implies that the Casimir element \eqref{ical} is an integral of motion of the system with the
zero conformal dimension.
Next, suppose that the Hamiltonian apart from itself and \eqref{ical}
possesses other integrals of motion.
Here we plan to study in detail the interrelation of the conformal symmetry
and these integrals.

Note that any constant of motion is a highest-weight vector of the conformal algebra \eqref{sl2},
since it is annihilated by the
Hamiltonian. Without any restriction, one can choose it to have a certain
conformal dimension (spin),
\begin{equation}
\label{hw}
\hat S_+ I_s= 0,
\qquad
\hat S_z I_s= s I_s,
\end{equation}
where we have introduced more conventional notations for the raising, lowering and diagonal
generators of SL(2,$\R$):
\begin{gather}
\label{Spmz}
S_+=H_0, \qquad
S_-=-K,\qquad
S_z=-\sfrac12 D,
\\
\label{commSS}
[ \hat S_+, \hat S_- ] =2\hat S_z, \qquad
[\hat S_z, \hat S_\pm]=\pm \hat S_\pm\,.
\end{gather}
The covariant basis for the conformal algebra reads
\begin{gather}
\label{Sxy}
 S_{x,y}=\sfrac12(S_+\pm S_-),
\\
\label{SS}
\{ S_x, S_y \}= -S_z, \qquad
\{ S_y, S_z \}= -S_x, \qquad
\{ S_z, S_x \}= S_y, \\
\label{S^2}
\mathbf{S}^2=\sfrac14 \ical=\sum_{\alpha=x,y,z} S_\alpha S^\alpha=-S_x^2+S_y^2-S_z^2,
\end{gather}
where the indices are raised and lowered by the metric
$g_{\alpha\beta}=\text{diag}(-1,1,-1)$.
The Casimir element $\mathbf{\hat S}^2$ of the related vector field algebra
\begin{equation}
\label{sl2-vf}
[\hat S_x, \hat S_y ]=  -\hat S_z, \qquad
[\hat S_y, \hat S_z ]=  -\hat S_x, \qquad
[\hat S_z, \hat S_x ]=  \hat S_y
\end{equation}
equals $-s(s{+}1)$ times the identity on an $sl(2,\R)$ representation of spin $s$ as given in \eqref{hw}.
It is important to distinguish between
the square of the conformal spin \eqref{sl2-vf}, as a second-order differential operator, and
the vector field $\hat\ical$ generated by the Casimir invariant \eqref{S^2}, as a first-order operator:
\be
\label{hat-S^2}
\mathbf{\hat S}^2=\sum_\alpha \hat S_\alpha \hat S^\alpha, \qquad
\hat\ical=\{\ical,\cdot\}=8\sum_\alpha  S_\alpha \hat S^\alpha.
\ee
Note that $\hat\ical$ does not preserve representations but acts as an intertwiner between them.

The descendants
\begin{equation}
\label{S^k}
I_{s,k}=(\hat S_-)^kI_s \qquad\textrm{for}\quad  k=0,1,2,\dots
\end{equation}
form the basic states of
the spin-$s$ representation of the conformal algebra \eqref{sl2-vf}.
For generic real values of $s$, there is an infinity of them.
For non-negative integer or half-integer values of $s$ however, the  state $I_{s,2s+1}$
either vanishes or it is another integral of motion for the conformal mechanics Hamiltonian.

If $I_{s,2s+1}=0$, then we deal with a finite-dimensional irreducible representation.
This includes the rational Calogero model, whose Liouville constants of motion are
polynomials of order $(2s)$ in the momenta.
Their multiplets are nonunitary and similar to the spin-$s$ representations of $su(2)$.

If $I_{s,2s+1}$  does not vanish, it is another integral of the conformal Hamiltonian,
since the spin raising operator $\hat S_+$ annihilates it, as is easy to check
using  the definition \eqref{hw} and commutation relations \eqref{commSS}.
As a highest-weight state, it generates another conformal representation, which forms
an invariant subspace. We thus encounter an indecomposable representation,
which is reducible but not fully reducible.

The time evolution of the observables \eqref{S^k} is given by
a $k$th order polynomial in time \cite{woj-sl2}, since the $(k{+}1)$th power of the evolution
operator $d/dt=\hat H_0=\hat S_+$ annihilates it. However, they can be used
to construct new integrals of motion different from $I_s$. This construction
is be done in terms of the representation theory of the conformal algebra,
as will be described below.

Denote by $(s)$ the $sl(2,\R)$ representation \eqref{S^k}, generated by the integral of motion $I_s$.
For two integrals $I_{s_1}$ and $I_{s_2}$ the products of the corresponding descendant states
form the product representation, which decomposes into a direct sum of representations:
\be
\label{sum-rules-infin}
(s_1)\otimes (s_2)=(s_1+s_2)\oplus (s_1+s_2-1)\oplus \ldots \oplus(s_1+s_2-k)\oplus \ldots\ .
\ee
For finite dimensional irreducible representations,
this series terminates at $(|s_1{-}s_2|)$, giving rise to the usual momentum
sum rule in quantum mechanics.
The highest-weight states of the $k$th multiplet in the decomposition \eqref{sum-rules-infin}
are also integrals of motion with conformal spins $s=s_1+s_2-k$, which we denote by $I^{(s_1,s_2)}_s$.
They can be calculated using the $sl(2,\R)$ Clebsch-Gordan coefficients.
However, it is easier to derive them directly with the commutation relation
\be
[\hat S_+,\hat S_-^l]=l\hat S_-^{l-1}(2\hat S_z-l+1)
\ee
and the highest-state conditions \eqref{hw}. Choosing a suitable normalization factor, we
define the new integrals of motion as
\be
\label{hw-mult}
I^{(s_1,s_2)}_{s_1+s_2-k}
=\sum_{l=0}^{k} (-1)^l\binom{k}{l} \frac{ \Gamma(2s_1-k+l+1)\Gamma(2s_2-l+1) }%
{ \Gamma(2s_1-k+1) \Gamma(2s_2-k+1)}  I_{s_1,k-l}I_{s_2,l}
\qquad\textrm{for}\quad k=0,1,2,\ldots\ .
\ee
For generic real values of $s_1$ or $s_2$, we obtain an infinity of these.
For integer or half-integer values of the spins however, there appears only a finite number,
limited by $k<2\textrm{min}(s_1,s_2)$.
In multi-particle models, the new integrals $I^{(s_1,s_2)}_s$ are quadratic in the descendants
and hence involve a double sum over the particle index, making them composite objects.

Let us consider particular cases of the composite integrals of motion \eqref{hw-mult}.
The case $k=0$ is uninteresting, since it merely yields the product $I_{s_1}I_{s_2}$.
For $k=1$ we have
\be
\label{int-woj}
I^{(s_1,s_2)}_{s_1+s_2-1}=2s_2 I_{s_1,1}I_{s_2}-2s_1 I_{s_1}I_{s_2,1}\ ,
\ee
which is a new integral of motion for any pair $I_{s_1}$ and $I_{s_2}$ \cite{woj83}.
If the first integral is the Hamiltonian, $I_{s_1}=I_1=S_+$,
then we simply obtain the bracket with the Casimir element \eqref{S^2},
\be
\label{ical-Is}
I^{(1,s) }_{s}=-4s S_z I_{s}-2 S_+\hat S_- I_s=-(4 S_z\hat S_z +2 S_+\hat S_- +2  S_-\hat S_+) I_s
=4\sum_\alpha S_\alpha\hat S^\alpha  I_s=\frac12\hat\ical  I_s\ .
\ee
In the last equation the second relation in \eqref{hat-S^2} is used.
For the $N$-particle Calogero system, the Casimir invariant $\ical$ of the conformal
algebra produces in this way the additional $N{-}1$ integrals of motion from the
Liouville integrals \cite{hkln,hlns}.
Another special case is $s_1=s_2$. Exchanging $l\leftrightarrow k-l$ in the sum \eqref{hw-mult},
one concludes that terms with $k$ odd cancel out, leaving only even values for $k$.

Rather than simply multiplying the descendants on the right-hand side of \eqref{hw-mult},
one may take their Poisson bracket instead, since they are phase-space functions.
The conformal tensor product decomposition is unaffected, and the Clebsch-Gordan coefficients
remains the same,
\be
\label{hw-poi}
\sum_{l=0}^{k} (-1)^l\binom{k}{l}
\frac{\Gamma(2s_1-k+l+1)\Gamma(2s_2-l+1) }{\Gamma(2s_1-k+1) \Gamma(2s_2-k+1)}
\{ I_{s_1,k-l}, I_{s_2,l}\}.
\ee
For the Calogero model, the Poisson bracket of two conformal representations
generated by   Liouville integrals $I_{s_1}$ and $I_{s_2}$,
was considered already in \cite{barucchi77}.
For the  standard Liouville integrals in the simplest cases the above formula
yields nothing new as we shall sketch in Section~5.
Therefore, we further discuss only the pointwise products \eqref{hw-mult}.

\section{SL(2,$\R$) structure of the integrals of motion: quantum case}

\noindent
In passing from the classical to the quantum model, we replace
\be
\{p_i,q_j\}=\delta_{ij} \qquad\longrightarrow\qquad \frac{i}{\hbar}[p_i,q_j]=\delta_{ij}\ .
\ee
The expressions \eqref{Calogero}, \eqref{Calogero-0} and \eqref{DK} for the Hamiltonians and conformal group
generators remain the same, except that symmetric (Weyl) ordering between momenta  and coordinates
must be used, which affects the dilatation
\be
\label{q-D}
D=\frac12\sum_{i=1}^N (p_iq_i+q_ip_i)=\sum_{i=1}^N p_iq_i+i\hbar N/2\ .
\ee
The hermitian generators obey the quantum commutation relations
\be
\label{q-sl2}
[H_0,D]=-2i\hbar H_0,
\qquad [K,D]=2i\hbar K,
\qquad [H_0,K]=-i\hbar D.
\ee
The expressions \eqref{Spmz} and \eqref{Sxy} for the invariant conformal
generators $S_{\pm,z,x,y}$ remain unchanged, while the quantum analogue of \eqref{SS} reads
\begin{gather}
\label{q-SS}
[S_\alpha, S_\beta ]= -i\hbar \epsilon_{\alpha\beta\gamma}  S^\gamma,
\\
  [  S_+, S_- ] =-2i\hbar S_z,
\qquad
[S_z, S_\pm]=\mp i\hbar  S_\pm.
\end{gather}
Note that, in contrast to the well known $su(2)$ raising and lowering operators,
the $sl(2,\R)$ operators $S_\pm$ are hermitian and thus not mutually conjugate.

The Weyl ordering becomes  essential in the Casimir element
\be
\label{q-S^2}
\ical=4\mathbf{S}^2,
\qquad
\mathbf{S}^2=-\frac12(S_+S_- + S_-S_+) - S_z^2
 = -S_-S_+ - S_z(S_z-i\hbar).
\ee

Any quantum observable $f$ defines an
infinitesimal evolution map given by the operator
\be
\label{q-hat-sl2}
\hat f= \frac{i}{\hbar} [f,\cdot],
\ee
which is the quantum analog of the classical vector field
\eqref{hat-f} and reduces to it in the semiclassical limit.
Again, the assignment $f\to\hat f$
is a Lie algebra homomorphism. In this way, we get a (not necessarily unitary)
representation of the conformal algebra on the space of quantum operators.
It was introduced and used for the construction of additional integrals
of the quantum Calogero system \cite{gonera98-2}, simplifying an earlier procedure
\cite{kuznetsov}.

In the adjoint action \eqref{q-hat-sl2}, the quantum commutation relations of
the conformal group generators coincide with their classical
commutators  \eqref{commSS} and \eqref{sl2-vf}.

As in the classical case, any spin-$s$ quantum integral of motion \eqref {hw}
generates a highest-weight representation \eqref{S^k} of the conformal
algebra. The product of two such representations is subject to the sum rule
\eqref{sum-rules-infin}, and the highest states \eqref{hw-mult} yield new
integrals of motion for the quantum conformal Hamiltonian. However, since
quantum physical observables are supposed to be hermitian,
the expressions for $I^{(s_1,s_2)}_{s_1+s_2-k}$ must be self-conjugate.
In order to achieve this,
it suffices to symmetrize the products of descendants in $I_s^{(s_1,s_2)}$, i.e. \footnote{
Other orderings, like Weyl ordering, are also possible, but will differ only
by contributions of lower-order integrals. Hence, the full set of quantum integrals
does not depend on the choice of ordering.}
\be
I_{s_1,k_1} I_{s_2,k_2} \qquad\longrightarrow\qquad
\frac12\bigl( I_{s_1,k_1} I_{s_2,k_2} +  I_{s_2,k_2} I_{s_1,k_1} \bigr)\, .
\ee
It appears that $I^{(s_1,s_2)}_{s_1+s_2-k}$ is simply the irreducible component
of the symmetrized tensor product $(I_{s_1}\otimes I_{s_2})_+$:
\be
\label{hw-pm}
I^{(s_1,s_2)}_{s_1+s_2-k}
=\sum_{l=0}^{k} (-1)^l\binom{k}{l} \frac{ \Gamma(2s_1{-}k{+}l)\Gamma(2s_2{-}l{+}1) }
{ \Gamma(2s_1{-}k{+}1)\Gamma(2s_2{-}k{+}1)}
\frac12 \bigl( I_{s_1,k-l}I_{s_2,l} + I_{s_2,l}I_{s_1,k-l} \bigr)\ .
\ee
In the classical limit $\hbar\to0$, it reduces to \eqref{hw-mult}.
The first nontrivial case corresponds to $k=1$:
\be
I^{(s_1,s_2)}_{s_1+s_2-1}=
s_2 \bigl( I_{s_1,1}I_{s_2} + I_{s_2}I_{s_1,1} \bigr) -
s_1 \bigl( I_{s_1}I_{s_2,1} + I_{s_2,1}I_{s_1} \bigr)\ ,
\ee
which is the quantum version of the classical integrals \eqref{int-woj} \cite{kuznetsov,gonera98-2}
and produces, for $s_1=1$ and $I_1=S_+$,
\be
I_s^{(1,s)} = \frac{i}{2\hbar}\bigl[ \ical,I_s \bigr]
\ee
as the simplest quantum integrals beyond the Liouville ones.

\section{Relating angular mechanics and confined conformal mechanics}

\noindent
The angular part $\ical$ of the Hamiltonian coincides with the Casimir element $\mathbf{S}^2$ of the
conformal algebra \eqref{S^2}. It defines a mechanical subsystem depending only on
the angular coordinates and momenta $u$, to which we refer as angular mechanics.
Its integrals of motion have been studied for general conformal mechanics~\cite{hkln,hlns} and,
in particular, for the Calogero model~\cite{hln}.

Motion in the angular subsystem \eqref{ical} is naturally bounded,
as it is for the harmonically confined conformal system~\eqref{Calogero}.
Confined integrable systems feature quantities which oscillate in time with a fixed frequency.
Examples are the angle variables, which, together with their canonically conjugated action variables,
the Liouville integrals, parametrize the phase space of the system.
In the case of commensurate frequencies, additional integrals exist and are expressed
completely via the angles~\cite{LL}. This property is known as superintegrability.
Note that the existence of such integrals does not require integrability:
two quantities oscillating with commensurate frequencies are sufficient.
In this section we study consequences of the existence of higher-order integrals in two
confined mechanical systems, namely in angular mechanics and in harmonically confined conformal mechanics.
For both we will construct quantities which oscillate in time with integral multiples of a basic frequency.
These frequencies are proportional to the spin projections in a finite-dimensional SL(2,$\R$) representation,
which is realized differently in the two systems.

Let us first focus on the angular mechanics case.
In order to construct the angular subsystem of a conformal mechanics model, it is
suitable to express the conformal generators \eqref{Spmz} and \eqref{hat-sl2} in
terms of angular coordinates and momenta, $u=(\theta_\alpha,p_{\theta_\alpha})$,
and of radial ones,
\be
r^2=\sum_i{q_i^2}, \qquad
rp_r=\sum_i{p_iq_i},
\ee
via
\begin{align}
\label{Sical}
& {S_+}=\frac{{p}^2_r}{2}+ \frac{\ical(u)}{2r^2},
&&{S_-}=-\frac{r^2}{2},
&&S_z=-\frac{p_r r}{2},
\\
\label{hat-Sical}
&\hat S_+ = p_r\part{r} + \frac{\ical}{r^3}\part{p_r}+
\frac{\hat\ical}{2r^2} ,
&&\hat S_-=r\part{p_r},
&&\hat S_z=\frac12 \left(p_r\part{p_r} -  r\part{r}\right).
\end{align}
Using \eqref{Sical} and \eqref{hat-Sical},
it is easy to see \cite{hlns} that the highest-weight condition \eqref{hw} is equivalent to
\begin{equation}
\label{hatI}
\hat \ical I_{s}=2(\hat S^R_+-\ical\hat S^R_-) I_{s},
\end{equation}
where we introduced the one-dimensional vector fields~\footnote{
The basis defined in our previous paper \cite{hlns}
differs from the current  one by the map $\hat S^R_\pm\to \hat S^R_\mp$ and
$\hat S^R_z\to -\hat S^R_z$. In this notation it becomes equivalent to the basis
of the previously defined conformal generators \eqref{hat-Sical}.}
\begin{equation}
\label{S-dual}
\hat S^R_+=-p_rr^2\frac{\partial}{\partial r},
\qquad
\hat S^R_-=\frac{1}{r}\frac{\partial}{\partial p_r},
\qquad
\hat S^R_z=\frac{1}{2}\left(r\frac{\partial}{\partial r}
+p_r\frac{\partial}{\partial p_r}\right),
\end{equation}
which form another $sl(2,\R)$ algebra.
Note that $\mathcal{\hat I}$ acts only on the angular variables
while the $\hat S_a$ feel just the radial dependence.
Therefore, \eqref{hatI} relates the angular dependence of $I_s$
to its radial one.

For vanishing angular part,
the above new generators are {\it dual\/} to the conformal generators \eqref{hat-Sical}:
\be
\label{duality}
\hat S_a|_{\ical=0}\;=\;R\hat S^R_aR,
\ee
where the duality map
\be
\label{R}
R: \quad r\to1/r, \quad p_r\to p_r, \quad u\to u
\ee
inverts the radial coordinate but leaves the angular ones unchanged.
Evidently, it is not a canonical transformation,
but the dual generators \eqref{S-dual} obey the same algebraic relations \eqref{commSS}
as the standard generators \eqref{hat-Sical} do.

For $2s$ being integer, any spin-$s$ integral, as defined in \eqref{hw},
can be decomposed into terms with a homogeneous radial dependence
separated from the angular one~\footnote{
In comparison to the definition of $f_{s,m}(u)$ in \cite{hlns},
we have multiplied a binomial factor and applied an index shift
$m\to m-s$. This makes the $so(3)$ properties more apparent and
simplifies further relations.},
\be
\label{decomp}
I_s(p_r,r,u) =\sum_{l=0}^{2s} f_{s,l}(u)\,\frac{p_r^{2s-l}}{r^l}.
\ee
The radial functions  $p_r^{2s-l}/r^l$  form a basis of the spin $s$-representation
of the $sl(2,\R)$ algebra \eqref{S-dual}, where $\hat S^R_z$ is diagonal.
The inversion \eqref{duality} maps them to the equivalent representation given by
the $S_a|_{\ical=0}$ acting on the polynomials $p_r^{2s-l}r^l$ of order~$2s$.
The conformal descendants of the integrals satisfy the following decomposition,
\be
\label{decomp-desc}
I_{s,k}(p_r,r,u) =\sum_{l=0}^{2s-k} \frac{(2s{-}l)!}{(2s{-}k{-}l)!}
\frac{p_r^{2s{-}k{-}l}}{r^{k+l}}f_{s,l}(u)
\qquad\textrm{for}\quad k=0,1,\ldots,2s.
\ee
In particular, the lowest descendant ($k=2s$) reduces to a single term
independent of the radial momentum, so its radial and angular dependencies factorize:
\be
\label{decomp-low}
I_{s,2s}(p_r, r, u) = (2s)!\ r^{-2s} f_{s,0}(u).
\ee
This is reminiscent of the wavefunctions of the quantum Calogero Hamiltonian \cite{flp}.
The transformation \eqref{decomp-desc} from $\{f_{s,2s-k}\}$ to $\{I_{s,k}\}$ is given by a
triangular matrix with diagonal elements $(k)!\ r^{-2s} $. Therefore, it is invertible,
and $f_{s,l}$ can be expressed in terms of $\{I_{s,2s},\ldots,I_{s,2s-l}\}$.

It is convenient to pass from the radial variables $1/r$ and $p_r$ to
the complex combinations
\be
\label{r-z}
z = \frac{1}{\sqrt{2}} \Bigl(p_r - \frac{i\sqrt{\ical}}{r}\Bigr),
\qquad
\bar z = \frac{1}{\sqrt{2}} \Bigl( p_r + \frac{i\sqrt{\ical}}{r}\Bigr),
\ee
As any linear map, it extends to the ($2s{+}1$)-dimensional space of polynomials of degree~$2s$,
\be
\label{basis-z}
z^{2s-l}\bar z^{l}
=(i\sqrt{\ical})^s\sum_{k=0}^{2s}\frac{p_r^{2s-k}}{r^k}\,(\widetilde{U}^s)_{kl},
\ee
where $\widetilde{U}^s$ is a $2s\times 2s$ matrix depending on $\sqrt{\ical}$.
Its elements can be derived from those of the fundamental representation,
$(\widetilde{U}^\frac12)_{kl}$, which are determined by~\eqref{r-z}.

In the new basis, the dualized spin operators take the form
\be
\label{zdz}
\hat S^R_z  = \frac12 \left( \bar z\part{z} + z\part{\bar z}\right), \qquad
\hat S^R_+ + \ical \hat S^R_- = i \sqrt{\ical}\left( z\part{\bar z}- \bar z\part{z}\right),\qquad
\hat S^R_+ - \ical \hat S^R_- = i \sqrt{\ical}\left( \bar z\part{\bar z}- z\part{ z}\right).
\ee
The decomposition \eqref{decomp} of a conformal mechanics integral in the new monomial basis
\eqref{basis-z} defines shifted angular harmonics $\tilde f_{s,l}(u)$ via
\be
\label{decomp-z}
I_s(z,\bar z,u) =\sum_{l=0}^{2s} \tilde f_{s,l}(u) \,z^{2s-l} \bar z^l
\qquad
\text{with}
\qquad
\tilde f_{s,l}=(i\sqrt{\ical})^{-s}\sum_{k=0}^{2s}  (\widetilde{U}^s)^{-1}_{lk} \, f_{s,k}.
\ee
We remark that our bases are not normalized.
The standard, normalized, $sl(2,\R)$ representation basis \cite{hlns}
is obtained by using
\cite{hlns}
\be
\label{basis-normal}
\sqrt{\binom{2s}{s{-}m}} \ \bar z^{s+m} z^{s-m}
\qquad\textrm{and}\qquad
\sqrt{\binom{2s}{s{-}m}} \ \frac{p_r^{s+m}}{ r^{s-m}},
\qquad\textrm{for}\quad
-s\le m\le s
\ee
together with related normalized angular harmonics
\be
\label{f-normal}
f_{sm}=\binom{2s}{s{-}m}^{-\frac12} f_{s,s-m}
\qquad\textrm{and}\qquad
\tilde f_{sm}=\binom{2s}{s{-}m}^{-\frac12} \tilde f_{s,s-m}.
\ee

According to the last equation in \eqref{zdz},
the action \eqref{hatI} of the angular Hamiltonian becomes diagonal in the new coordinates.
Hence, its action on the decomposition  \eqref{decomp-z} implies that, for $\ical>0$,
the shifted harmonics oscillate with the frequencies equal to integral multiples
of $\sqrt{\ical}$:
\be
\label{eigen}
\hat\ical {\tilde f}_{sm}=-2mi\sqrt{\ical} {\tilde f}_{sm}
\qquad\longrightarrow\qquad
\tilde f_{sm}(t)=e^{-2mi\sqrt{\ical}(t-t_0)  }{\tilde f}_{sm}(t_0).
\ee
The basic frequency $\sqrt{\ical}$ is, of course, a constant of motion.

The map \eqref{basis-z} defined by the matrix $\widetilde U^s$
can be expressed $s$-independently in terms of the conformal generators,
\be
\label{U-R}
\widetilde{U}\equiv\widetilde{U}^s\big(\sqrt{\ical}\,\big)
= \big(i\sqrt{\ical}\,\big)^{-\hat S^R_z}e^{\frac{\pi}2 \hat S^R_y}.
\ee
In the fundamental representation $s=1/2$,
the operators $\hat S^R_y$ and $\hat S^R_z$ are represented in terms of Pauli matrices
as  $\sfrac i2\sigma_y$ and $\sfrac12\sigma_z$, respectively,
but the expression naturally extends to the polynomial spin-$s$ representation of the
conformal algebra \eqref{zdz}. On the basis \eqref{basis-z} or \eqref{basis-normal}, it
evidently reduces to the matrix $\widetilde{U}^s$ from \eqref{decomp-z}. From \eqref{U-R} it is clear
that the latter can be expressed in terms of Wigner's small $d$-matrix,
which is explicitly done in Appendix \ref{app:A}.
With the help of \eqref{Um'm} and \eqref{f-normal}, the relation between the normalized
original and shifted angular harmonics take the following form,
\begin{gather}
\label{f-ftilde}
\tilde f_{sm}=\sum_{m'=-s}^s d_{mm'}^s(\pi/2) \big(i\sqrt{\ical}\,\big)^{m'-s} f_{sm'},
\\
\label{ftilde-f}
f_{sm'}=\sum_{m=-s}^s d_{mm'}^s(\pi/2)   (i\sqrt{\ical})^{s-m'} \tilde f_{sm}.
\end{gather}

Using the definition \eqref{Sxy} and  the relation
$q^{-\hat S_z} \hat S_\pm q^{\hat S_z} = q^{\mp 1}\hat S_\pm$
for $q=i\sqrt{\ical}$, which is a direct consequence of the commutation relations
\eqref{commSS}, the adjoint action of the operator \eqref{U-R} on the
generators of the conformal algebra  \eqref{S-dual} or \eqref{zdz} can be
calculated:
\be
\label{conj-ical}
\hat S^R_z= \widetilde{U} \,\hat S^R_x\, \widetilde{U}^{-1}, \qquad
\hat S^R_++\ical\hat S^R_- = 2i\sqrt{\ical}\, \widetilde{U} \,\hat S^R_y\, \widetilde{U}^{-1}, \qquad
\hat S^R_+-\ical\hat S^R_- = -2i\sqrt{\ical}\, \widetilde{U} \,\hat S^R_z\, \widetilde{U}^{-1}.
\ee
We emphasize again that the operator \eqref{U-R}  is not canonical since the vector fields
$\hat S^R_z$ and $\hat S^R_y$  are not Hamiltonian.
The expression \eqref{conj-ical}  is, in general, complex and multi-valued.
When the potential is positive, as is the case in Calogero models, the angular
part is strictly positive and the operator \eqref{U-R}
is complex but single-valued.  In any case, all square roots will cancel in
the final expressions for the constants of motion.

\medskip

The second part of this section deals with conformal mechanics in an external harmonic potential.
We shall see that, again, the integrals of motion are derived from the descendants $I_{s,k}$
with the help of a Wigner rotation, similar to the angular mechanics case above.
Adding a harmonic confining potential is a deformation compatible with the conformal symmetry:
\be
\label{Hw}
H_\omega=H_0+\omega^2 K = S_+- \omega^2 S_-,
\qquad
\hat H_\omega = \hat S_+- \omega^2 \hat S_-.
\ee
The last relation has the same structure as in \eqref{hatI},
and it is mapped to the latter expression under the substitution
\be
\label{omega-ical}
\hat S_a\to \hat S^R_a,
\qquad
\omega \to \sqrt{\ical}.
\ee
However, while \eqref{hatI} is valid only for the integrals of motion $I_s$ of $H_0$,
\eqref{Hw} is an operator identity.
Using this formal analogy between \eqref{hatI} and \eqref{Hw},
one can recycle the previous subsection to express the constants of motion for $H_\omega$
in terms of the integrals for $H_0$.
This procedure generalizes a construction previously applied to the Calogero model \cite{gonera99-2}.

Using \eqref{conj-ical} and \eqref{U-R} and the
correspondence \eqref{omega-ical}, it is easy to see that the operator
\be
\label{U}
U =U(\omega)= (i\omega)^{-\hat S_z}e^ {\frac{\pi}2 \hat S_y}
\ee
links the Hamiltonian with harmonic potential to the diagonal conformal generator,
\be
\label{H-Sz}
U:\ -2i\omega S_z\ \mapsto\ H_\omega, \qquad
\hat H_\omega=-2i\omega U \hat S_z U^{-1}.
\ee
The operator $U$ defines a complex-valued nonlocal map which, in contrast to
its counterpart $\widetilde{U}$, is canonical. Nevertheless,
its action on the space spanned by the descendants \eqref{S^k} is given by an
SL(2,$\C$) representation matrix.
The SL(2,$\C$) transformation \eqref{U}
mapping $S_z$ to $H_\omega$
is determined only
up to an overall $z$ rotation from the right~\footnote{
Indeed, \cite{gonera99-2} used instead the following operator for the Calogero model:
$\omega^{-\hat S_z}e^ {\frac{i\pi}2 \hat S_x}= U i^{\hat S_z}.$}.
Our choice of \eqref{U} is fixed by the transformation rules
\be
\label{conj-Hw}
U:\ S_x\ \mapsto\  S_z
\quad\textrm{and}\quad
2iS_y\ \mapsto\ \omega^{-1}S_++\omega S_-.
\ee

The operator \eqref{U} diagonalizes also the basis \eqref{S^k} of the spin-$s$ representation
comprising the descendants of the spin-$s$ integral of conformal mechanics.
In complete analogy with \eqref{basis-z}, we define
\be
\label{basis-I}
\tilde I_{s,l}:=
(i\omega)^s\sum_{k=0}^{2s} I_{s,k}\,(U^s)_{kl},
\ee
where the matrix elements $(U^s)_{kl}$ depend on $\omega$.
Applying the second relation in \eqref{H-Sz} to these shifted basis states, we get
\be
\label{H-I}
\hat H_\omega\tilde I_{s,l}
=-2im\omega \tilde I_{s,l}
\qquad
\text{with}
\qquad
m=s{-}l.
\ee
Due to these eigenvalue relations, the ``harmonics'' \eqref{I-tildeI}
oscillate in time with integer frequencies proportional to the spin projection value,
\be
\label{I-osc}
\tilde I_{sm}(t)=e^{-2im\omega (t-t_0)} \tilde I_{sm}(t_0).
\ee

Here and in the following, we use the standard basis for $sl(2,\R)$ representations
by introducing
\be
\label{I-norm}
I_{sm}=\sqrt{\sfrac{(s{+}m)!}{(2s)!(s{-}m)!}}\, I_{s,s-m},
\qquad
\tilde I_{sm}=\sqrt{\sfrac{(s{+}m)!}{(2s)!(s{-}m)!}}\, \tilde I_{s,s-m}
\qquad\textrm{with}\quad -s\le m\le s,
\ee
which is distinguished from the previous basis by omitting the comma between indices.
In this basis, the decomposition \eqref{decomp-desc} into radial and angular parts reads
\be
\label{decomp-m}
I_{sm}=\sum_{m'=-m}^s\sqrt{\binom{s{+}m}{m{+}m'}
\binom{s{+}m'}{m{+}m'}}\,\frac{p_r^{m+m'}}{r^{2s-m-m'}}\,f_{sm'},
\ee
where we applied the same index nomenclature to the $f$s given by \eqref{f-normal}.

As was mentioned before, the transformation $U$ is canonical and preserves the Poisson brackets,
\be
\label{canonic}
U:\ \{ I_{s_1m_1}, I_{s_2m_2}\}\ \mapsto\ \{\tilde I_{s_1m_1},\tilde I_{s_2m_2}\}.
\ee
In particular, if some $ \tilde I_{sm}$ are in involution, then the corresponding $ I_{sm}$
are in involution, too.

The matrix form of the action of the shift operators \eqref{U} and \eqref{U-R} acquires the following form,
\begin{gather}
\label{I-tildeI}
\tilde I_{sm}=\sum_{m'=-s}^s d_{m'm}^s(\pi/2) (i\omega)^{s-m'} I_{sm'},
\\
\label{tildeI-I}
I_{sm'}=\sum_{m=-s}^s d_{m'm}^s(\pi/2)  (i\omega)^{m'-s}\tilde I_{sm}.
\end{gather}
These formulae are analogous to \eqref{f-ftilde} and \eqref{ftilde-f}. Actually,
the transformation \eqref{I-tildeI} is equivalent to the transformation
\eqref{basis-z}, which is the inverse transpose of \eqref{f-ftilde},
according to the definition \eqref{decomp-z}.

Finally, we would like to take advantage of the structural analogy of the two models and
directly relate the corresponding shifted harmonics \eqref{f-ftilde} and \eqref{I-tildeI}.
To this end, we first substitute \eqref{ftilde-f} into \eqref{decomp-m} and then insert
the resulting expression for $I_{sm}$ into \eqref{I-tildeI}. Ultimately, we arrive at
\be
\label{tilde-If}
\tilde I_{sm}=\sideset{}{'}\sum_{m',m_1,m_2} c^s_{m_1m_2,mm'}(i\omega)^{s-m_2}
(i\sqrt{\ical})^{s-m_1}\frac{p_r^{m_2+m_1}}{r^{2s-m_2-m_1}}\,\tilde f_{sm'},
\ee
which expresses the harmonic functions of the confined conformal mechanics
in terms of the  harmonics of the related angular system.
Here, the prime over the sum restricts the indices by the condition $m_1+m_2\ge 0$,
so that all arguments of the binomial coefficients in
\be
\label{c}
c^s_{m_1m_2,mm'}=\sqrt{\binom{s{+}m_1}{m_1{+}m_2}\binom{s{+}m_2}{m_1{+}m_2}}\
d_{m_1 m}^s(\pi/2)\ d_{m'm_2}^s(\pi/2)
\ee
are positive.

\newpage

\section{The rational Calogero model with harmonic potential}

\noindent
In this section we specialize to the Calogero model and employ the well known matrix-model description.
Firstly, for the unconfined Calogero Hamiltonian $H_0$~\eqref{Calogero-0}, we work out the explicit form
of the integrals of motion \eqref{hw-mult} composed of the conformal descendants of two standard
Liouville integrals. Secondly, for the confined Calogero Hamiltonian $H_\omega$~\eqref{Calogero},
we present a simple expression for the oscillating observable $\tilde I_{sm}$ of Section~4.
Thirdly, we act with the Hamiltonian vector field $\hat \ical$ related to the angular Hamiltonian on the
$N$ standard Liouville integrals of $H_\omega$ and obtain the additional $N{-}1$ integrals for this model.
The functional independence of all $2N{-}1$ integrals is proven explicitly, demonstrating that they comprise
a complete system. A similar property is already known for $H_0$ \cite{hkln,hlns}.

The Hamiltonians \eqref{Calogero} and \eqref{Calogero-0} can be obtained by SU$(N)$
reduction, respectively, from the hermitian matrix models \cite{kazhdan,perelomov81,polychronakos}
\footnote{
We apply the same notation $H_0$ and $H_{\omega}$ for the Calogero systems and the related matrix models.
All other notations like $I_{sm}$ or $\tilde I_{sm}$ are preserved also.}
\bea
\label{P2+Q2}
\label{matrix}
H_\omega=\frac12\tr{P^2}+\frac{\omega^2}{2}\tr{Q^2}
\qquad\textrm{and}\qquad
H_0=\frac12\tr{P^2}.
\eea
Here, $P$ and $Q$ are hermitian matrices containing the canonical momenta
and coordinates, subject to
\be
\label{commPQ}
\{P_{ij},Q_{i'j'}\}=\delta_{ij'}\delta_{ji'}\,.
\ee
The round brackets denote the SU($N$) trace,
\be
\label{tr}
(X):=\text{tr} X.
\ee
The matrix Hamiltonians \eqref{P2+Q2} describe a homogeneous $N^2$-dimensional oscillator
and a free particle in $\R^{N^2}$, respectively.
In the matrix-model reduction, the Calogero coupling $g$ is recovered by the gauge-fixing relation
\be
[P,Q]=-ig(1-e\otimes e) \quad \text{with}  \quad e=(1,1,\ldots,1).
\ee
Up to SU($N$) transformations, the coordinates and Lax matrix of the Calogero model \eqref{Calogero-0}
provide the solution of above equation \cite{kazhdan,perelomov81}:
\be
\label{Lax}
Q_{ij}=\delta_{ij}q_i, \qquad
P_{ij}=\delta_{ij}p_i+(1-\delta_{ij})\frac{ig}{q_i-q_j}.
\ee
As a result, the matrix Hamiltonians \eqref{matrix} are reduced to the corresponding Calogero
models \eqref{Calogero} and \eqref{Calogero-0}.

The generators \eqref{Calogero-0} and \eqref{DK} of the conformal algebra
\eqref{sl2} acquire the following form in the matrix-model representation:
\be
\label{DKH}
S_z=-\frac12\tr{PQ},
\qquad
S_-=-\frac12\tr{Q^2},
\qquad
S_+=\frac12\tr{P^2}.
\ee
Furthermore, the standard Liouville integrals and their descendants \eqref{S^k}
of the Calogero Hamiltonian $H_0$ are obtained from the reduction
of \cite{barucchi77}:
\be
\label{Is}
I_s=\tr{P^{2s}} \qquad\textrm{and}\qquad
I_{s,l}=\frac{(2s)!}{(2s{-}l)!}\tr{P^{2s-l}Q^l}_\text{sym},
\ee
respectively. Here, $s\le N/2$, and the index `sym' means symmetrization over all orderings of $P,Q$
matrices inside the trace, e.g.
\be
\tr{P^nQ}_\text{sym}=\tr{P^nQ},
\qquad
\tr{P^2Q^2}_\text{sym}=\frac23 \tr{P^2Q^2}+\frac13\tr{PQPQ}.
\ee
The symmetrized traces can be computed by means of the generating function
\be
\label{gen}
\tr{(P+vQ)^{2s}}=\sum_{l=0}^{2s} \binom{2s}{l}\tr{P^{2s-l}Q^{l}}_\text{sym}v^l
=\sum_{l=0}^{2s}\frac{v^l}{l!}I_{s,l},
\ee
which can be considered as an extension of Newton's binomial formula.

Using \eqref{Is}, we can write down the integral \eqref{hw-mult} in terms of symmetrized traces,
\be
\label{hw-pq}
I^{(s_1,s_2)}_{s_1+s_2-k}
=\frac{(2s_1)!(2s_2)!}{(2s_1{-}k)!(2s_2{-}k)!}\sum_{l=0}^k (-1)^l \binom{k}{l}
\tr{P^{2s_1-k+l} Q^{k-l} }_\text{sym}  \tr{ P^{2s_2-l} Q^l}_\text{sym} .
\ee
The coefficient in front of the sum is not essential and can be ignored.
Substituting  $s_1=1$, $s_2=n/2$ and $k=1$ in above equation, we arrive at the integral
proportional to $\{\ical,I_\frac n2\}$, as was shown in \eqref{ical-Is},
\be
\label{additional}
I_\frac n2^{(1,\frac n2)}\ \sim\ \hat\ical I_\frac n2\ \sim\
\tr{P Q } \tr{ P^n} - \tr{P^2} \tr{ P^{n-1}Q} .
\ee
For $n=2$ it vanishes, but the remaining $N{-}1$ integrals together with the Liouville ones
$\tr{P^n}$, $1\le n\le N$,
constitute a complete system of constants of motion of $H_0$. The functional independence of this set
can be seen from the free-particle limit $g=0$ where, according to \eqref{Lax}, the matrices
$P$, $Q$ are diagonal. This property reveals the role of the angular part:
it generates the full system of integrals for $H_0$ by acting on the Liouville ones.
Below we will show that this role of $\ical$ extends to the confined system $H_\omega$ as well.
For the particular case of $s_1=s_2=n/2$ and $k=2$ in \eqref{hw-pq}, we find the integral
\be
\label{hw-pq-2}
I^{(\frac{n}{2},\frac{n}{2})}_{n-2}
\ \sim\ \tr{P^{n-2}Q^{2}}_\text{sym} \tr{P^n} -\tr{P^{n-1}Q}^2.
\ee
At $n=2$ we recover the Casimir element of \eqref{DKH} describing the angular mechanics.

Before passing to the confined model $H_\omega$, we briefly consider the issue of the $H_0$
integrals \eqref{hw-poi} obtained by taking Poisson brackets.
Let us take Liouville integrals \eqref{Is} for $I_{s_1}$ and $I_{s_2}$.
In the $k=0$ special case, \eqref{hw-poi} reduces to $\{I_{s_1},I_{s_2}\}=0$,
since $I_{s_1}$ and $I_{s_2}$ are in involution.
For $k=1$ one finds the Liouville integral $I_{s_1+s_2-1}$. The $k=2$ case
vanishes again, as can be calculated using \eqref{commPQ} and \eqref{Is}.
In general, the relations \eqref{Lax} imply that
the Poisson brackets in \eqref{hw-poi} evaluate to
\be
\{ I_{s_1,k_1}, I_{s_2,k_2}\}\ \sim\ (s_1k_2-s_2k_1) I_{s_1+s_2-1,k_1+k_2-1}+\ldots\ ,
\ee
where the remaining terms are of order $O(g)$ and thus vanish in the free-particle
limit. Their structure is more complicated: for higher spins they may contain, besides traces,
also mean values $\langle e| \ldots  |e\rangle$ of products of $P,Q$ matrices.
In total, one obtains the whole infinite algebra of observables of the Calogero model \cite{avan95}.

For studying the Calogero system in an external harmonic potential, it is most suitable
to employ the creation and annihilation combinations
\be
\label{A-PQ}
A^\pm=\frac{1}{\sqrt{2\omega}}P \pm i\sqrt{\frac{\omega}{2}} Q.
\ee
In terms of these, the Hamiltonian $H_\omega$ reads \cite{polychronakos}
\be
\label{A+A}
H_\omega=\omega \tr{A^+A^-}
\qquad\textrm{with}\qquad
\{A^-_{ij},A^+_{i'j'}\}=i\delta_{ij'}\delta_{ji'}.
\ee
The matrix variables $A^\pm$ oscillate in time with frequency $\omega$:
\be
\label{iwt}
\dot{A}^\pm=\{H_\omega,A^\pm\} = \pm i\omega A^\pm
\qquad\longrightarrow\qquad
A^\pm(t)=e^{\pm i\omega (t-t_0)} A^\pm(t_0).
\ee

Using the canonical brackets in \eqref{A+A} and the expressions \eqref{DKH}
for the SL(2,$\R$) generators, one can calculate the action of the transformation
\eqref{U} on the phase-space variables of the matrix model:
\be
\label{UPQ}
UP=e^{-i\frac\pi4} A^-, \qquad
UQ= e^{i\frac\pi4} A^+.
\ee
Recall that this transformation maps the diagonal generator $S_z$
to the Hamiltonian $H_\omega$  according to \eqref{H-Sz}.
It turns into the analogous transformation $\widetilde{U}$ given by \eqref{r-z}
upon substituting
\be
\label{ztoA}
(P,Q)\to (p_r,r^{-1}), \qquad (\sqrt{\omega}A^-,\sqrt{\omega}A^+)\to (z,\bar z),
\qquad
\omega\to\sqrt {\ical}.
\ee
In this context, the analogy between confined between confined Calogero model and
related angular system becomes more transparent in the matrix model description.
It expands also to the homogeneous polynomials,
which form spin-$s$ representations of conformal algebras
According to \eqref{I-norm} and \eqref{Is}, the matrix form of original and shifter normalised
states are:
\begin{gather}
\label{I-PQ}
I_{sm}=\sqrt{\sbinom{2s}{s-m}}\tr{P^{s+m}Q^{s-m}}_\text{sym},
\\
\label{I-A}
\tilde I_{sm}=\sqrt{\sbinom{2s}{s-m}}\,\omega^s\tr{(A^-)^{s+m}(A^+)^{s-m}}_\text{sym}.
\end{gather}
According to \eqref{I-osc},  $\tilde I_{sm}$ oscillates with the frequency $-2m\omega$, which also can be
seen from \eqref{iwt}.

In fact, the trace of any product of $A^\pm$ matrices,
$$
\tr{A^{\sigma_1}\dots A^{\sigma_n}} \qquad \text{with} \qquad \sigma_i\in\{+,-\}
$$
oscillates with integer frequency equal to $\omega \sum_{i}\sigma_i$.
Any product of such observables with a vanishing total sum of the $\sigma_i$ will be
an integral of motion of the Calogero system (\ref{Calogero}).
The number of such integrals is significantly higher than $2N{-}1$,
but they are not independent.

The Liouville integrals of the Hamiltonian $H_\omega$
can be extracted from this general set using
either the Lax-pair method or the symmetries of the original matrix
model \eqref{P2+Q2} or \eqref{A+A}. We recall the second way described
in the review \cite{polychronakos}.
In terms of the hermitian left-multiplication generator or Lax matrix $A^+A^-$,
one has
\be
\label{In}
\tilde I_n=\tr{(A^+A^-)^n}.
\ee
The related flow is not symplectic, because the left (or
right) multiplication does not preserve the Poisson brackets in (\ref{A+A}),
as the adjoint action does.
The matrix elements of $A^+A^-$ obey the U($N$) commutation relations,
and the Liouville integrals \eqref{In} can be identified with the
Casimir elements of that group \cite{polychronakos}.

The Liouville integrals $\tilde I_n$ are unsymmetrized analogs of the integrals $\tilde I_{s=n\,m=0}$
\eqref{I-A} with integer spin $s=n\in\NN$. They form another set of Liouville integrals \cite{gonera99-2}.
Up to a numerical factor, the integrals $\tilde I_{n0}$ and $\tilde I_n$ coincide in their
term of highest power in the momenta, corresponding to the free-particle limit $g=0$.
Note that, according to \eqref{canonic}, the descendants $I_{n0}$ also are in involution.
The first integral is proportional to the Hamiltonian and
corresponds to the central U(1) part,
\be
\tilde I_1= \tr{A^+A^-}=\omega^{-1}H_\omega=\omega^{-1}H_0+\omega K.
\ee
The remaining  two bilinear traces of \eqref{A-PQ} are also related to the conformal generators.
Using   their matrix form \eqref{DKH} and \eqref{Spmz}, we obtain:
\be
\tr{A^+A^+}=\omega^{-1}H_0-\omega K- iD, \qquad
\tr{A^-A^-}=\omega^{-1}H_0 -\omega K+ iD.
\ee
>From these equations it is easy to express  the angular Hamiltonian,
described by the Casimir element \eqref{ical}
of the conformal algebra,  in terms of $A^\pm$ matrices. It has a rather simple form:
\be
\ical=\tr{A^+A^-}^2-\tr{A^+A^+}\tr{A^-A^-}.
\ee
Now recall that  although the conformal generators \eqref{sl2} are not symmetries
of the Hamiltonian $H_\omega$, their Casimir element is conserved:
$\dot{\mathcal{I}}=\{H_\omega,\mathcal{I}\}=0$.
The invariant $\mathcal{I}$ is not a Liouville integral since does
not commute with the whole set (\ref{In}). Its Poisson brackets with
the Liouville integrals give rise to additional integrals like for $H_0$:
\be
\label{Jn}
J_n=\{\mathcal{I}, \tilde I_n\}
= 2in\left[ \tr{A^-A^-}\tr{A^+A^+(A^+A^-)^{n-1}}-\tr{A^+A^+}\tr{A^-A^-(A^+A^-)^{n-1}}\right].
\ee
Since $J_1=0$, we find $2N{-}1$ integrals $(\tilde I_1,\dots,\tilde I_N,J_2,\dots J_N)$,
which appear to form a complete set of integrals for $H_\omega$.
Their functional independence is shown in Appendix~\ref{app:B}.
This confirms the superintegrability of the Calogero model in the harmonic potential.
The construction of the additional integrals $J_n$ is similar to the one for the free
Calogero Hamiltonian described by equation \eqref{additional} above.

\begin{acknowledgments}
\noindent
We are grateful to A.\ Nersessian for simulating discussions.
This work has been supported by the \hbox{VolkswagenStiftung}
under the grant  no.\ 86 260.
Further support was given
by the Armenian State Committee of Science, 
grants no. 13RF-018 (T.H., D.K.), 13-1C114 (T.H.),  13-1C132 (D.K.), 
and by ANSEF grants no. 3501 (T.H.) and 3122 (D.K.).
\end{acknowledgments}

\pagebreak

\pagebreak

\appendix
\numberwithin{equation}{section}

\section{}
\label{app:A}

\noindent
In this appendix we calculate  the transforation matrix \eqref{basis-z},
then express it in terms of the Wigner's small $d$-matrix.

The explicit expression for the inverse $\widetilde{U}^s$-matrix from \eqref{decomp-z} can be obtained
by substitution of the inverse transformations \eqref{r-z} into the decomposition \eqref{decomp},
with subsequent comparison of  the obtained angular coefficients with $f_{s,k}$ in \eqref{decomp}.
As a result, we get:
\be
\label{U-inv}
(\widetilde{U}^s)^{-1}_{lk}=  \big(i\sqrt{\ical}\big)^{s-k}b_{lk}^s,
\ee
where the above $b$-matrix satisfies  the following relation:
\be
\label{monomials}
2^{-s}(z+\bar z)^{2s-k} (\bar z- z)^k
=\sum_{l=0}^{2s} b_{lk}^s z^{2s-l}\bar z^{l}.
\ee
Its explicit form can be calculated using the Newton's binomial formula:
\be
\label{b-matrix}
b_{lk}^s=
\sum_{t=\max(0,l+k-2s)}^{\min(k,l)}  \frac{(-1)^{k+t}}{ 2^s}\binom{k}{t}\binom{2s{-}k}{l{-}t}
= \sum_{t=\max(0,l+k-2s)}^{\min(k,l)} i
\frac{(-1)^{k+t}}{2^s}\frac{k!(2s{-}k)!}{t!(k{-}t)!(l{-}t)!(2s{-}l{-}k{+}t)! }.
\ee
Note that the sum over $t$ in \eqref{b-matrix} is taken over all nonnegative
values of the four factorials in the denominator.

Then it is easy to see that the map $z\to (z{+}\bar z)/\sqrt{2}$, $\bar z\to (\bar z{-} z)/\sqrt{2}$
is orthogonal. Therefore, the matrix \eqref{b-matrix}
is orthogonal too:
\be
\sum_i b^s_{li} b^s_{ki}=\delta_{lk}.
\ee
Hence, from \eqref{U-inv} we obtain:
\be
\label{U-dir}
(\widetilde{U}^s)_{kl}= \big(i\sqrt{\ical}\big)^{k-s}\,b_{kl}^s.
\ee

Next, following \cite{hlns}, we express the matrix $b^s_{kl}$  in terms of Wigner's small $d$-matrix, which
describes an SU(2) rotation around the $y$ axis:
$d_{m'm}^s(\beta)=\langle sm'| \exp(-\beta \hat S^R_y)| sm\rangle$.
Comparing the last expression in \eqref{b-matrix} with
the formula (2) in \S4.3 of \cite{ang-momentum}
we get:
\be
\label{blk}
b^s_{lk}=
\sqrt{\sfrac{k! (2s-k)!}{l!(2s-l)!}}\,
d_{s-l\,s-k}^s(\pi/ 2).
\ee

Finally, using  \eqref{U-dir} and \eqref{U-inv}, we obtain the explicit forms of the
$\widetilde{U}^s$-matrix and its inverse:
\begin{gather}
\label{Um'm}
(\widetilde{U}^s)_{s-m'\,s-m}=
\sqrt{\sfrac{(s-m)! (s+m)!}{(s-m')!(s+m')!}}\,
d^s_{m'm }(\pi/2)\,\big(i\sqrt{\ical} \big)^{s-m'},
\\
(\widetilde{U}^s)^{-1}_{s-m\,s-m'}=
\sqrt{\sfrac{(s-m')! (s+m')!}{(s-m)!(s+m)!}}\,
d^s_{mm' }(\pi/2)\,\big(i\sqrt{\ical} \big)^{m'-s}.
\end{gather}

\section{}
\label{app:B}

\noindent
In this appendix we prove that the $2N{-}1$ integrals $(I_1,\dots,I_N,J_2,\dots
J_N)$ defined by \eqref{In} and \eqref{Jn}, are functionally independent.
Here, we omit the tilde on $\tilde I_i$,
since here we deal only with $H_\omega$ and its integrals.
It suffices to prove their independence for the free-particle limit
$g\to 0$, since this projects to the highest-order term in momenta for the
polynomials $I_n$ and $J_n$. In this limit, the reduced $P$ and $Q$ matrices
given by \eqref{Lax} are diagonal:
\be
A=\textrm{diag}(a_1,\dots,a_N)+O(g), \qquad
a_i=\frac{p_i}{\sqrt{2\omega}} - i\sqrt{\frac{\omega}{2}} q_i
=: \sqrt{\rho_i}e^{-i\varphi/2}.
\ee
For the integrals we thus have
\be
I_n=\sum_{i=1}^N\rho_i^{n}
\qquad\textrm{and}\qquad
J_n=4n\sum_{i,j=1}^N\rho_i\rho_j^{n}\sin(\varphi_i-\varphi_j).
\ee

The functional independence of this set of integrals is equivalent to
the nondegeneracy of the Jacobian matrix
\be
\everymath{\displaystyle} \frac{\partial(I_1,\dots,I_N,J_2,\dots
J_N)}{\partial(\rho_1,\dots,\rho_N,\varphi_2,\dots,\varphi_{N})}
=\left(
\begin{array}{cc}
\frac{\partial(I_1,\dots,I_N)}{\partial(\rho_1,\dots,\rho_N)} &
\frac{\partial(I_1,\dots,I_N)}{\partial(\varphi_2,\dots,\varphi_N)}
\\[12pt]
\frac{\partial(J_2,\dots,J_N)}{\partial(\rho_1,\dots,\rho_N)} &
\frac{\partial(J_2,\dots,J_N)}{\partial(\varphi_2,\dots,\varphi_N)}
\end{array}\right).
\ee
Due to the obvious relations
\be
\label{J-phi}
\frac{\partial I_n}{\partial \varphi_k}=0,
\qquad
\frac{\partial I_n}{\partial \rho_k}
=n\rho_k^{n-1},
\qquad
\frac{\partial J_n}{\partial \varphi_k}
=\sum_i(\rho_k\rho_i^{n}-\rho_i\rho_k^{n})\cos(\varphi_k-\varphi_i),
\ee
the Jacobi matrix has block-triangular form, so its determinant is
given by the product
\be
\everymath{\displaystyle} \left|
\frac{\partial(I_1,\dots,I_N)}{\partial(\rho_1,\dots,\rho_N)}
\right|
\;
\left|
\frac{\partial(J_2,\dots,J_N)}{\partial(\varphi_2,\dots,\varphi_N)}
\right|.
\ee
The first term is proportional to the Vandermonde determinant
\be
\label{vandermonde}
\everymath{\displaystyle} \left|
  \frac{\partial(I_1,\dots,I_N)}{\partial(\rho_1,\dots,\rho_N)}
\right|
=N!\prod_{1\le i<j\le N }(\rho_j-\rho_i),
\ee
while the  second one is more complicated.
It can be presented as a
product of two rectangular matrices with dimensions  $(N{-}1)\times N$ and
$N\times(N{-}1)$. Indeed, according to  (\ref{J-phi}),
\be
\label{a-ik}
\frac{\partial J_n}{\partial\varphi_k}
=\sum_{i=1}^N\rho _i^{n-1}B_{ik}
\qquad\textrm{with}\qquad
B_{ik}=b_{ik}-\delta_{ik}\sum_{l=1}^Nb_{lk}
\quad\textrm{and}\quad
b_{ik}=\rho_i\rho_k\cos(\varphi_i-\varphi_k).
\ee
It is possible to express the matrix equation (\ref{a-ik}) in terms of square
$N\times N$ matrices with an additional row and column via
\be
\label{wij}
\left(
\ba{cccc}
1&0&\ldots&0\\
\rho_1&{\partial J_2}/{\partial \varphi_2}&\ldots&\partial J_2/\partial \varphi_N\\
\vdots&\vdots&\ddots&\vdots\\
\rho_1^{N-1}&{\partial J_N}/{\partial \varphi_2}&\ldots& {\partial J_N}/{\partial \varphi_N}
\ea
\right)
=
\left(
\ba{cccc}1&1&\ldots&1\\
\rho_1&\rho_2&\ldots&\rho_N\\
\vdots&\vdots&\ddots&\vdots\\
\rho^{N-1}_1&\rho^{N-1}_2&\ldots&\rho^{N-1}_N
\ea
\right)
\cdot
\left(
\ba{cccc}1&B_{12}&\ldots&B_{1N}\\
0&B_{22}&\ldots&B_{2N}\\
\vdots&\vdots&\ddots&\vdots\\
0&B_{N2}&\ldots&B_{NN}
\ea
\right).
\ee
Indeed, its restriction to the $\partial J/\partial \varphi$ block coincides with
(\ref{a-ik}). The  first column of this matrix product is
obvious, while the zeros in the first row  appear due to the property
$\sum_{i= 1}^N B_{ij}=0$, which follows from the definition (\ref{a-ik}).

This relation factorizes out the Vandermonde determinant from the
 Jacobian $\frac{\partial (J_2\dots J_N)}{\partial(\varphi_2\dots \varphi_N)}$.
The remaining matrix can be further reduced by extracting the diagonal matrix
$B_{ik}=\tilde B_{ik}\rho_k$, where $\tilde B$ is obtained
by the substitution
\be
b_{ik}\to\tilde b_{ik}=\rho_i\cos(\varphi_i-\varphi_k)
\ee
in the definition of $B$ in (\ref{a-ik}). Together with (\ref{wij}) this implies
\be
\left|
\frac{\partial(J_2,\dots,J_N)}{\partial(\varphi_2,\dots,\varphi_N)}
\right|
=\prod_{i=2}^N \rho_i \prod_{1\le i<j\le N}(\rho_j-\rho_i) \;\;M_{11},
\ee
where $M_{ij}$ denotes a minor of the matrix $\tilde B$.

In the simplest case of equal phases $\varphi_i=\varphi$, this
determinant is given by characteristic polynomial of the rank-one matrix
$\tilde b_{ij}=\rho_j$ with the value $\rho=\sum _{i=1}^N\rho_i$ for
the eigenvalue variable, which can be easy calculated:
\be
M_{11}=\det(\tilde B_{ij}-\rho\delta_{ij})=\rho^{N-1}\rho_1.
\ee
Therefore, the Jacobian vanishes at special points only.
The matrix $\tilde B$ generically depends
analytically on the arguments $\varphi_i$, so its determinant can
vanish only at a set of measure zero in the space of $\varphi_i$.


\begin{thebibliography}{99}

\bibitem{calogero69}
F.~Calogero,
J. Math. Phys. {\bf 10} (1969)
\href{http://dx.doi.org/10.1063/1.1664820}{2191};
{\sl ibid.} {\bf 12} (1971)
\href{http://dx.doi.org/10.1063/1.1665604}{419}.

\bibitem{moser}
J.~Moser,
Adv. Math. {\bf 16} (1975)
\href{http://dx.doi.org/10.1016/0001-8708(75)90151-6}{197}.

\bibitem{trig-Cal}
B.~Sutherland,
Phys. Rev. A {\bf 4} (1971)
\href{http://dx.doi.org/10.1103/PhysRevA.4.2019}{2019};
Phys. Rev. A {\bf 5} (1972)
\href{http://dx.doi.org/10.1103/PhysRevA.5.1372}{1372}.

\bibitem{spin-Cal}
J.~Gibbons and T.~Hermsen,
Physica D {\bf 11} (1984)
\href{http://dx.doi.org/10.1016/0167-2789(84)90015-0}{337};
S.~Wojciechowski,
Phys. Lett. A {\bf 111} (1985)
\href{http://dx.doi.org/10.1016/0375-9601(85)90432-3}{101}.

\bibitem{super-Cal}
D.Z.~Freedman and P.F.~Mende,
Nucl. Phys.  B {\bf 344} (1990)
\href{http://dx.doi.org/10.1016/0550-3213(90)90364-J}{317}.

S.~Fedoruk, E.~Ivanov and O.~Lechtenfeld,
Phys. Rev.  D {\bf 79} (2009) 105015,
\href{http://arxiv.org/abs/0812.4276}{arXiv:0812.4276}.

\bibitem{algebra}
J.~Wolfes,
J. Math. Phys.  {\bf 15} (1974)
\href{http://dx.doi.org/10.1063/1.1666826}{1420}.

F.~Calogero and C.~Marchioro,
J. Math.  Phys.  {\bf 15} (1974)
\href{http://dx.doi.org/10.1063/1.1666827}{1425}.

M.A.~Olshanetsky and A.M.~Perelomov,
Lett. Math. Phys.  {\bf 2} (1977)
\href{http://dx.doi.org/10.1007/BF00420664}{7}.

\bibitem{frac-stat}
A.P.~Polychronakos,
Nucl. Phys. B {\bf 324} (1989)
\href{http://dx.doi.org/10.1016/0550-3213(89)90522-1}{597}.

\bibitem{frac-Hall}
H.~Azuma and S.~Iso,
Phys. Lett. B  {\bf 331} (1994)
\href{http://dx.doi.org/10.1016/0370-2693(94)90949-0}{107}.

\bibitem{haldane-rvb}
F.D.M.~Haldane,
Phys. Rev. Lett.  {\bf 60} (1988)
\href{http://dx.doi.org/10.1103/PhysRevLett.60.635}{635}.

\bibitem{gibbons}
G.W.~Gibbons and P.K.~Townsend,
Phys. Lett. B {\bf 454} (1999) 187,
\href{http://arxiv.org/abs/hep-th/9812034}{hep-th/9812034}.

\bibitem{woj83}
S.~Wojciechowski,
Phys. Lett. A {\bf 95} (1983)
\href{http://dx.doi.org/10.1016/0375-9601(83)90018-X}{279}.

\bibitem{kuznetsov}
V.B.~Kuznetsov,
Phys. Lett. A {\bf 218} (1996) 212-222,
\href{http://arxiv.org/abs/solv-int/9509001}{arXiv:solv-int/9509001}.

\bibitem{gonera98-2}
C.~Gonera,
Phys. Lett. A {\bf 237} (1998)
\href{http://dx.doi.org/10.1016/S0375-9601(98)00903-7}{ 365}.

\bibitem{adler}
M.~Adler,
Comm. Math. Phys. {\bf 55} (1977)
\href{http://dx.doi.org/10.1007/BF01614548}{195}.

\bibitem{perelomov81}
M.A.~Olshanetsky and A.M.~Perelomov,
Phys.  Rept.  {\bf 71}, (1981)
\href{http://dx.doi.org/10.1016/0370-1573(81)90023-5}{313};
Phys. Rept.  {\bf 94} (1983)
\href{http://dx.doi.org/10.1016/0370-1573(83)90018-2}{313}.

\bibitem{gonera99-2}
C.~Gonera and P.~Kosinski,
Acta Phys. Polon. B {\bf 30} (1999) 907,
\href{http://arxiv.org/abs/hep-th/9810255}{hep-th/9810255}.

\bibitem{gonera98-1}
C.~Gonera,
J. Phys. A {\bf 31} (1998)
\href{http://dx.doi.org/10.1088/0305-4470/31/19/012}{4465}.

\bibitem{feher10}
V.~Ayadi and L.~Feher,
Phys. Lett. A {\bf 374} (2010) 1913-1916,
\href{http://arxiv.org/abs/0909.2753}{arXiv:0909.2753}.

\bibitem{feher12}
V.~Ayadi, L.~Feher and T.F.~Gorbe,
J. Geom. Symmetry Phys. {\bf 27} (2012) 27-44,
\href{http://arxiv.org/abs/1209.1314}{arXiv:1209.1314}.

\bibitem{woj-sl2}
S.~Wojciechowski,
Phys. Lett. A {\bf 64} (1978)
\href{http://dx.doi.org/10.1016/0375-9601(77)90359-0}{273}.

\bibitem{gonera99-1}
T.~Brzezinski, C.~Gonera and P.~Maslanka,
Phys. Lett. A {\bf 254} (1999) 185,
\href{http://arxiv.org/abs/hep-th/9810176}{hep-th/9810176}.

\bibitem{lech06}
A.~Galajinsky, O.~Lechtenfeld and K.~Polovnikov,
Phys. Lett. B {\bf 643} (2006) 221-227,
\href{http://arxiv.org/abs/hep-th/0607215}{arXiv:0607215}.

\bibitem{gonera00}
T.~Brzezinski, C.~Gonera, P.~Kosinski and P.~Maslanka,
Phys. Lett. A {\bf 268} (2000) 178,
\href{http://arxiv.org/abs/hep-th/9912068}{hep-th/9912068}.

\bibitem{hkln}
T.~Hakobyan, S.~Krivonos, O.~Lechtenfeld and A.~Nersessian,
Phys. Lett.  A {\bf 374} (2010) 801,
\href{http://arxiv.org/abs/0908.3290}{arXiv:0908.3290}.

\bibitem{cuboct}
T.~Hakobyan, A.~Nersessian and V.~Yeghikyan,
J.  Phys. A  {\bf 42}  (2009) 205206,
\href{http://arxiv.org/abs/0808.0430}{arXiv:0808.0430}.

\bibitem{hlns}
T.~Hakobyan, O.~Lechtenfeld, A.~Nersessian and A.~Saghatelian,
J. Phys. A  {\bf 44} (2011) 055205,
\href{http://arxiv.org/abs/1008.2912}{arXiv:1008.2912}.

\bibitem{hln}
T.~Hakobyan, O.~Lechtenfeld and A.~Nersessian,
Nucl. Phys. B  {\bf 858} (2012) 250,
\href{http://arxiv.org/abs/1110.5352}{arXiv:1110.5352}.

\bibitem{flp}
M.~Feigin, O.~Lechtenfeld and A.~Polychronakos,
JHEP {\bf 1307} (2013) 162,
\href{http://arxiv.org/abs/1305.5841}{arXiv:1305.5841}.

\bibitem{conf-mech}
V.~de Alfaro, S.~Fubini and G.~Furlan,
Nuovo Cim. A {\bf 34} (1976)
\href{http://dx.doi.org/10.1007/BF02785666}{569}.

\bibitem{kumar98}
P.~Claus, M.~Derix, R.~Kallosh, J.~Kumar, P.~Townsend and A.~van Proeyen,\\
Phys. Rev. Lett. {\bf 81} (1998) 4553,
\href{http://arxiv.org/abs/hep-th/9804177}{hep-th/9804177}.

\bibitem{gal-ners13}
A.~Galajinsky and A.~Nersessian,
JHEP {\bf 1111} (2011) 135,
\href{http://arxiv.org/abs/1108.3394}{arXiv:1108.3394};

A.~Galajinsky, A.~Nersessian and A.~Saghatelian,
JHEP {\bf 1306} (2013) 002,
\href{http://arxiv.org/abs/1303.4901}{arXiv:1303.4901}.

\bibitem{galaj-bh}
A.~Galajinsky,
Phys. Rev. D {\bf 78} (2008) 044014,
\href{http://arxiv.org/abs/0806.1629}{arXiv:0806.1629};
JHEP {\bf 1011} (2010) 126,
\href{http://arxiv.org/abs/1009.2341}{arXiv:1009.2341};

A.~Galajinsky and K.~Orekhov,
Nucl. Phys. B {\bf 850} 339-348,
\href{http://arxiv.org/abs/1103.1047}{arXiv:1103.1047};

S.~Bellucci and S.~Krivonos,
JHEP {\bf 1110} (2011) 014,
\href{http://arxiv.org/abs/1106.4453}{arXiv:1106.4453}.

\bibitem{barucchi77}
G.~Barucchi and T.~Regge,
J. Math. Phys. {\bf 18} (1977)
\href{http://dx.doi.org/10.1063/1.523384}{1149}.

\bibitem{LL}
L.D.~Landau and E.M.~Lifshitz,
\emph{Mechanics}, Moscow, ``Nauka'' (1988).

\bibitem{polychronakos}
A.P.~Polychronakos,
J. Phys. A  {\bf 39} (2006) 12793,
\href{http://arxiv.org/abs/hep-th/0607033}{hep-th/0607033}.

\bibitem{kazhdan}
D.~Kazhdan, B.~Kostant and S.~Sternberg,
Comm. Pure Appl. Math. {\bf 31} (1978)
\href{http://dx.doi.org/10.1002/cpa.3160310405}{481}.

\bibitem{avan95}
J.~Avan and E.~Billey,
Phys. Lett. A \textbf{198}  (1995) 183,
\href{http://arxiv.org/abs/hep-th/9404040}{hep-th/9404040}.

\bibitem{ang-momentum}
D.A.~Varshalovich, A.N.~Moskalev and V.K.~Khersonskii,
\emph{Quantum theory of angular momentum},\\
World Scientific Publishing, 1988,
ISBN: 978-9971-5-0107-5.

\end{thebibliography}
\end{document}